\documentclass[twocolumn,showpacs,amsmath,amssymb,aps,pre,reprint,longbibliography]{revtex4-2}
\usepackage[pdftex]{graphicx}
\usepackage{amsmath}
\DeclareMathOperator{\sign}{sgn}
\begin{document}

\title{
Motility Induced Phase Separation and Frustration in Active Matter Swarmalators    
} 
\author{
B. Adorj{\' a}ni$^{1}$, A. Lib{\' a}l$^{1}$, C. Reichhardt$^{2}$, and C. J. O. Reichhardt$^{2}$ 
} 
\affiliation{
$^{1}$ Mathematics and Computer Science Department, Babe{\c s}-Bolyai University, Cluj 400084, Romania\\
$^{2}$ Theoretical Division and Center for Nonlinear Studies,
Los Alamos National Laboratory, Los Alamos, New Mexico 87545, USA}

\date{\today}
\begin{abstract}
We introduce a system of active matter swarmalators composed of elastically interacting run-and-tumble active disks with an internal phase $\phi_i$. The disks experience an additional attractive or repulsive force with neighboring disks depending upon their relative difference in $\phi_i$. In the absence of the internal phase, the system forms a Motility-Induced Phase Separated (MIPS) state, but when the swarmalator interactions are present, a wide variety of other active phases appear depending upon whether the interaction is attractive or repulsive and whether the particles act to synchronize or anti-synchronize their internal phase values. These include a gas-free gel regime, arrested clusters, a labyrinthine state, a regular MIPS state, a frustrated MIPS state for attractive anti-synchronization, and a superlattice MIPS state for attractive synchronization.
\end{abstract}
\maketitle

\section{Introduction}
Active matter systems contain self-motile particles
and can be found in soft matter \cite{Marchetti13,Bechinger16,Gompper20},
biological \cite{Lauga09,DiLeonardo10,Bhattacharjee19,Tan22,Hall23},
social \cite{Helbing01,Castellano09}, 
active topological \cite{Bowick22,Shankar22}, 
and robotic  \cite{Rubenstein14,Wang21,BenZion23} contexts.
The simplest models of active matter consist of
interacting elastic disks undergoing a run-and-tumble motion 
or driven Brownian diffusion. Even in the absence of any attractive forces,
such motion produces what is known as Motility Induced Phase Separation (MIPS)
when the activity level and the disk density are sufficiently large
\cite{Fily12,Redner13,Palacci13,Buttinoni13,Bialke15,Cates15,Ginot18,Paoluzzi22,Omar23}. For monodisperse particles, the dense clusters in the MIPS
state have triangular order and coexist with a lower density gas.
An open question is whether there can be other types of MIPS regimes that
have alternative ordering within the clusters. Also unknown is whether
inclusion of more complex particle-particle interactions
can lead to a comprehensive way to connect different
MIPS regimes within one model. 

In many soft and condensed matter systems that have competing
or multi-length scale interactions,
additional larger scale patterning can arise
\cite{Seul95,Malescio03,Glaser07,Chen11a,Khalil12,Pine12}.
Frustration effects can occur when not all of the constraints
in the system can be satisfied simultaneously,
as is the case for proton ordering in water ice \cite{Pauling35},
frustrated colloidal systems \cite{Han08,OrtizAmbriz16},
spin ice \cite{Ramirez03},
artificial spin ice systems \cite{Nisoli13,Skjaervo19},
and certain types of metamaterials \cite{Guo23}.
There are potentially many active systems where
additional competing interactions could be added that would
create additional regimes or frustration effects that might
interfere with the crystalline ordering in the dense phase of MIPS.

\begin{figure*}[ht]
\includegraphics[width=0.9\textwidth]{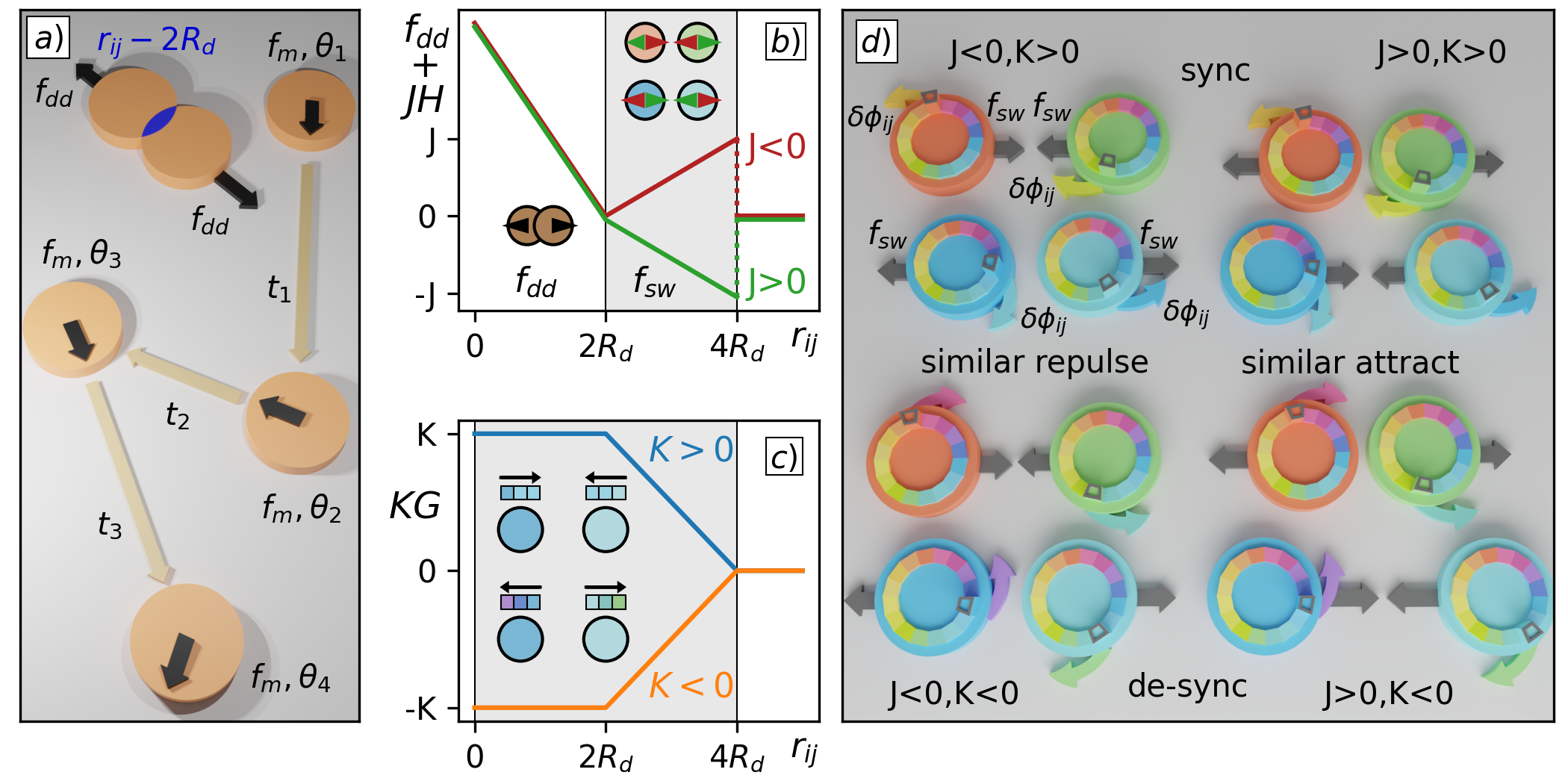}
\caption{Schematics of the system and the disk-disk interactions.
(a) Colliding disks experience a steric repulsion force $f_{\rm dd}$. Each disk
also undergoes run-and-tumble motion with a motor force of $f_{\rm m}$ directed
along $\theta_i$ for a time $t_i$, and
new values of $\theta_i$ and $t_i$ are selected during each instantaneous
tumbling event.
(b) $f_{\rm dd}+JH$, the disk-disk forces and the function $JH(r_{ij})$, 
vs the distance $r_{ij}$ between disks. Only steric disk repulsion acts
when $r_{ij}\leq 2R_d$. The value of $JH(r_{ij})$
becomes nonzero when
$2R_d < r_{ij} \leq 4R_d$; it is negative when $J>0$ (green) and positive
when $J<0$ (red). The left inset shows the steric interactions (brown disks).
The upper center inset shows swarmalator interactions between disks of different
phase, which is repulsive for $J>0$ (green arrows)
and attractive for $J<0$ (red arrows).
The lower center inset shows the swarmalator interactions when the disks
have similar phases, which is attractive for $J>0$ (green arrows) and repulsive
for $J<0$ (red arrows).
(c) The function $KG(r_{ij})$ controlling the phase evolution.
When $K>0$ (blue), the phase of neighboring disks becomes synchronized
over time,
as illustrated by the boxes in the top inset that illustrate
how the phase changes as a function of time.
When $K<0$ (orange), disk phases become desynchronized
over time, as shown by
the boxes in the lower inset.
(d) Summary of swarmalator forces for the four possible nonzero combinations
of $J$ and $K$. Disk color indicates the value of the phase, and the possible
phase values appear as a ring inside each disk. Grey arrows show the
orientation of $f_{\rm sw}$ for each combination, while the curved colored arrows
indicate the direction in which the phase will evolve as a result of the
interaction. For each combination, the phases of the disk pairs are different
in the upper row and similar in the lower row.
}
\label{fig:1}
\end{figure*}

Here we consider a model of run-and-tumble active disks that exhibit a MIPS phase for sufficiently high activity and density.
We give each disk an internal phase $\phi_i$ and
introduce an additional attractive or repulsive force
between neighboring disks $i$ and $j$ based on the swarmalator rules for
internal phase differences $\delta\phi_{ij}$ \cite{OKeefe17}.
This internal phase
evolves as a function of time and also changes due to
interactions with neighboring disks located within a radius
$2r_{\rm sw}$.
A schematic illustration of the model appears in Fig.~\ref{fig:1}.
The parameter $J$ controls the strength and sign of the
translational force
exerted along the line connecting two disks $i$ and $j$ depending
on the magnitude of $\delta\phi_{ij}$.
A second parameter $K$ determines how rapidly the value of $\phi_i$
changes in response to $\delta\phi_{ij}$, with positive values of $K$
resulting in a net reduction of $\delta\phi_{ij}$ and negative $K$
values causing $\delta\phi_{ij}$ to increase.
With these additional interactions,
we find that a wide variety of active matter regimes can be realized,
including distinctive MIPS states that revert back to the
ordinary MIPS state in the limit $J=0$, $K=0$.
For $J>0$, the MIPS state contains a hexagonal superlattice structure of disks with similar phase values surrounded by a gas of misaligned disks.
We also observe a multi-phase gel state at $K = 0$ and a frustrated liquid at high $K$ values where the superlattice is unable to form.
We map the evolution of these distinct states as a function of activity and density and find that MIPS is enhanced for the combination $K<0$ and $J>0$, but is suppressed for other combinations.

Our model could be realized using robotic swarms with steric interactions
and swarmalator rules \cite{Barcis19,Ceron23}.
Similar interactions could arise for biological systems
of motile cells of different types in which individual cells are
attracted to one cell type but repelled by a different cell type
\cite{Jiao14}.
There are also a variety of soft matter systems
that can have particle-specific or competing
attractive and repulsive interactions with each other
or that have multiple length scales in their interactions,
which can create pattern forming states
\cite{Seul95,Jagla98,Malescio03,Glaser07,Reichhardt10,Pine12}.
It should also be feasible to create competing interactions using
light activated colloids with feedback loops
\cite{Pince16,Lavergne19,Bauerle20}.

\section{Model}
We consider a two-dimensional system of size
$S_x=160$ and $S_y=160$ with periodic boundary
conditions in the $x$ and $y$ directions containing
$N = 4000$ circular disks of radius of $R_d = 1.0$.
The disk density is given by $\rho=N\pi R_d^2/(S_xS_y)$.
As shown in Fig.~\ref{fig:1}(a), disk $i$ 
interacts sterically at short
range with other disks according to
${\bf f}_{\rm dd} = f_{\rm dd}{\bf \hat{r}}_{ij}$
with $f_{\rm dd}=\sum_j^N k (2R_{d}-r_{ij})\Theta(2R_d-r_{ij})$,
where $r_{ij}=|{\bf r}_j-{\bf r}_i|$ is the distance between the
centers of disks $i$ and $j$, ${\hat {\bf r}}_{ij}=({\bf r}_j-{\bf r}_i)/r_{ij}$,
$\Theta$ is the Heaviside step function,
and $k = 20.0$ is the elastic constant.
Run-and-tumble motion is produced by a motor force ${\bf f}_m$ of
magnitude $f_m=1$
that is directed along $\theta_i \in [0, 2\pi)$ during a time
$t_i \in [\tau, 2\tau]$. 
At the end of each running event, an instantaneous tumbling event occurs
in which new values of $\theta_i$ and $t_i$ are chosen
randomly from the allowed intervals.

Each particle is endowed with an internal phase $\phi_i \in [0, 2\pi)$ that
is not coupled to $\theta_i$.
The internal phase undergoes a slow thermal diffusion that is independent of the
surroundings of the disk.
The thermal force $F_i^T$ is implemented using Langevin kicks with the
properties $\langle F^T_i\rangle = 0$ and
$\langle F^T_i(t) F^T_i(t^\prime)\rangle=2k_BT\delta_{ij}\delta(t-t^\prime)$.
Here we use a small thermal force magnitude of $F^T=0.001.$
The value of $\phi_i$ is also modified by a forcing term
of strength $K$ that is dependent
on the relative phases $\delta_{ij}=(\phi_j-\phi_i)$
of
neighboring disks within a radius of $r_{ij} \leq 2r_{\rm sw}$
where $r_{\rm sw}=2R_d$. This gives:
\begin{equation}
  \frac{d\phi_i}{dt} = \sum_{j}^N K \sin(\delta \phi_{ij})G(r_{ij}) + F^T_i
\end{equation}
where $G(r_{ij})=1$ for $r_{ij}\leq r_{\rm sw}$,
$G(r_{ij})=(2r_{\rm sw}-r_{ij})/r_{\rm sw}$ for $r_{\rm sw} < r_{ij} \leq 2r_{\rm sw}$,
and $G(r_{ij})=0$ for $r_{ij}>2r_{\rm sw}$.
The form of $G(r_{ij})$ is illustrated in
Fig.~\ref{fig:1}(c).
When $K>0$, the phases of neighboring disks tend to become synchronized, while
when $K<0$, the phases tend to become anti-synchronized.

The positions of the disk centers are obtained by integrating the
following equation of motion:
\begin{equation}
\eta\frac{d{\bf r}_i}{dt} = {\bf f}_{\rm m} + {\bf f}_{\rm dd} + {\bf f}_{\rm sw}  
\end{equation}
where we set the damping coefficient to $\eta=1$.
The swarmalator interaction force is given by
${\bf f}_{\rm sw}=f_{\rm sw}{\bf \hat{r}}_{ij}$ with
\begin{equation}
f_{\rm sw} = -\sum_j^N J \cos(\delta \phi_{ij})H(r_{ij}) \ ,
\end{equation}
where $H(r_{ij})=0$ for $r_{ij} \leq r_{\rm sw}$ or $r_{ij}>2r_{\rm sw}$
and $H(r_{ij})=(r_{ij}-r_{\rm sw})/r_{\rm sw}$ for
$r_{\rm sw} < r_{ij} \leq 2r_{\rm sw}$.
As shown in Fig.~\ref{fig:1}(b), disks with similar phase repel each other
when $J<0$ and attract each other when $J>0$.
When $J=0$, the internal phase has no impact on the disk dynamics and
the model reverts to the normal MIPS behavior found in the absence of
swarmalator interactions. Figure~\ref{fig:1}(d) summarizes the four
possible interaction regimes for nonzero combinations of $J$ and $K$.

We numerically integrate equations (1) and (2)
using a simulation time step of $dt = 0.001$.
To initialize the sample,
we place all disks in randomly chosen non-overlapping positions
and set the initial values of $\theta_i$, $\phi_i$, and $t_i$ to random
values chosen from the allowed range of each quantity.
We allow the system to evolve for $2\times 10^6$ simulation time steps before
collecting data during the next $3\times 10^6$ simulation time steps.

\section{Results}

\begin{figure}
\includegraphics[width=\columnwidth]{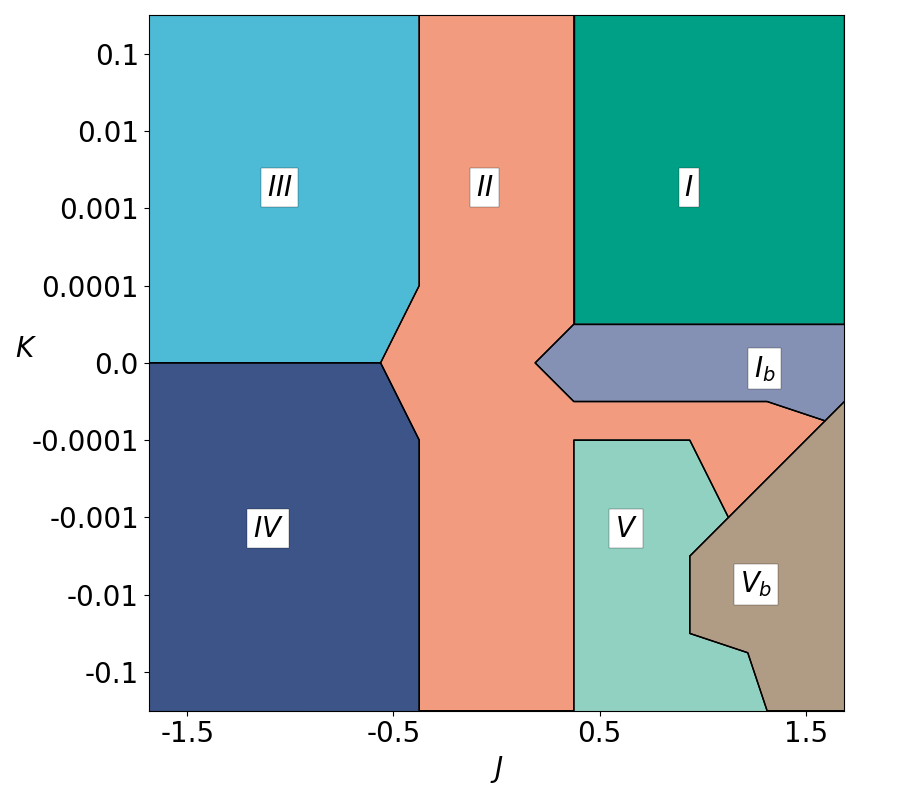}
\caption{
Dynamic phase diagram as a function of
$K$, the synchronization factor, and $J$, the swarmalator force strength,
for an active disk system with a density of $\rho = 0.49$
at $\tau=1.5 \times 10^5$. 
Region I with $K>0$ and $J>0$
is the active gel where all of the disks synchronize to the same phase,
as shown in Fig.~\ref{fig:3}(a). 
Region II is the ordinary
MIPS phase for $J = 0$, illustrated in Fig.~\ref{fig:3}(h). 
Region III for $J<0$ and $K>0$ is the active labyrinthine state
shown in Fig.~\ref{fig:3}(b)
where the disks synchronize to the same phase.
In region IV for $K<0$ and $J<0$ we find the frustrated MIPS phase
illustrated in Fig.~\ref{fig:3}(c,e) with stripe ordering in the clusters,
while region V for $K<0$ and $J>0$ is a superlattice MIPS phase
as shown in Fig.~\ref{fig:3}(d,f).
Region I$_b$ is the arrested gel phase shown in Fig.~\ref{fig:4}(a),
and region V$_b$ is a frustrated cluster fluid,
illustrated in Fig.~\ref{fig:4}(b).}
\label{fig:2}
\end{figure}

\begin{figure*}[ht]
\includegraphics[width=\textwidth]{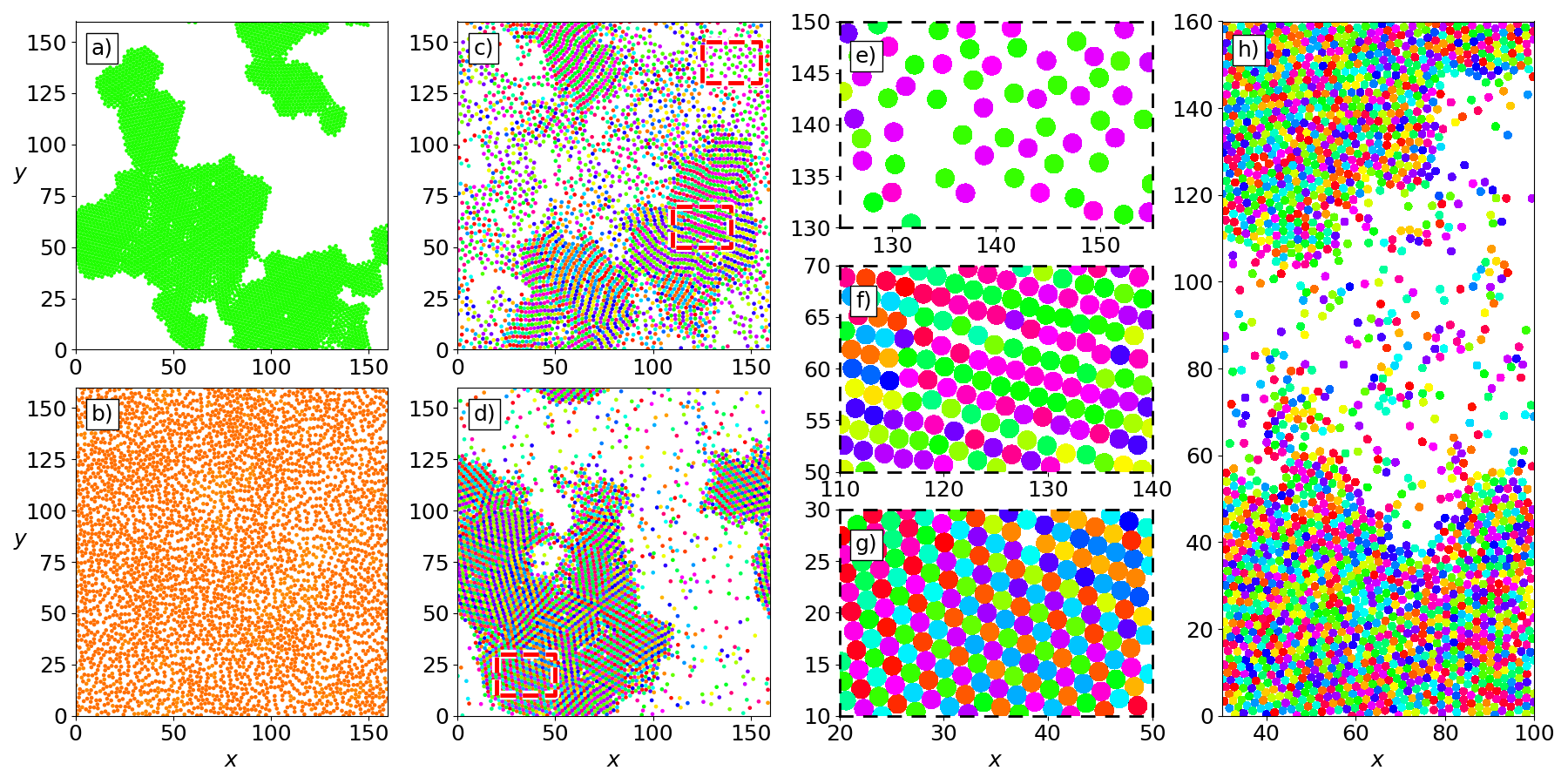}
\caption{
Illustrations of the disk locations (circles) and internal phases
(indicated by color) for
the dynamic phases described in Fig.~\ref{fig:2} in a system with $\rho=0.49$
and $\tau=1.5 \times 10^5$.
(a) Region I, the gel state where all disks have the same value of $\phi_i$,
at $J = 1.0$ and $K = 0.1$.
(b) Region III, the fluid labyrinthine state where all disks have the same
value of $\phi_i$, at $J = -1$ and $K = 0.001$.
(c) Region IV, the frustrated MIPS state with stripe ordering of $\phi_i$ inside
the clusters, at $J = -1.5$ and $K = -0.1$.
(d) Region V, the superlattice MIPS state, at $J =1.0$ and $K = -0.1$.
(e) A blowup of the upper red dashed box in panel (c) showing the
phase correlated fluid portion of the frustrated MIPS state.
(f) A blowup of the lower red dashed box in panel (c) showing stripe
ordering in the dense portion of the frustrated MIPS state.
(g) A blowup of the red dashed box in panel (d) showing a detail
of the superlattice ordering in the superlattice MIPS state.
(h) Region II, the MIPS state with no phase ordering,
at $J = 0$ and $K = 0$.
Videos of these phases are available in the supplemental material.
}
\label{fig:3}
\end{figure*}

\begin{figure}[ht]
\includegraphics[width=\columnwidth]{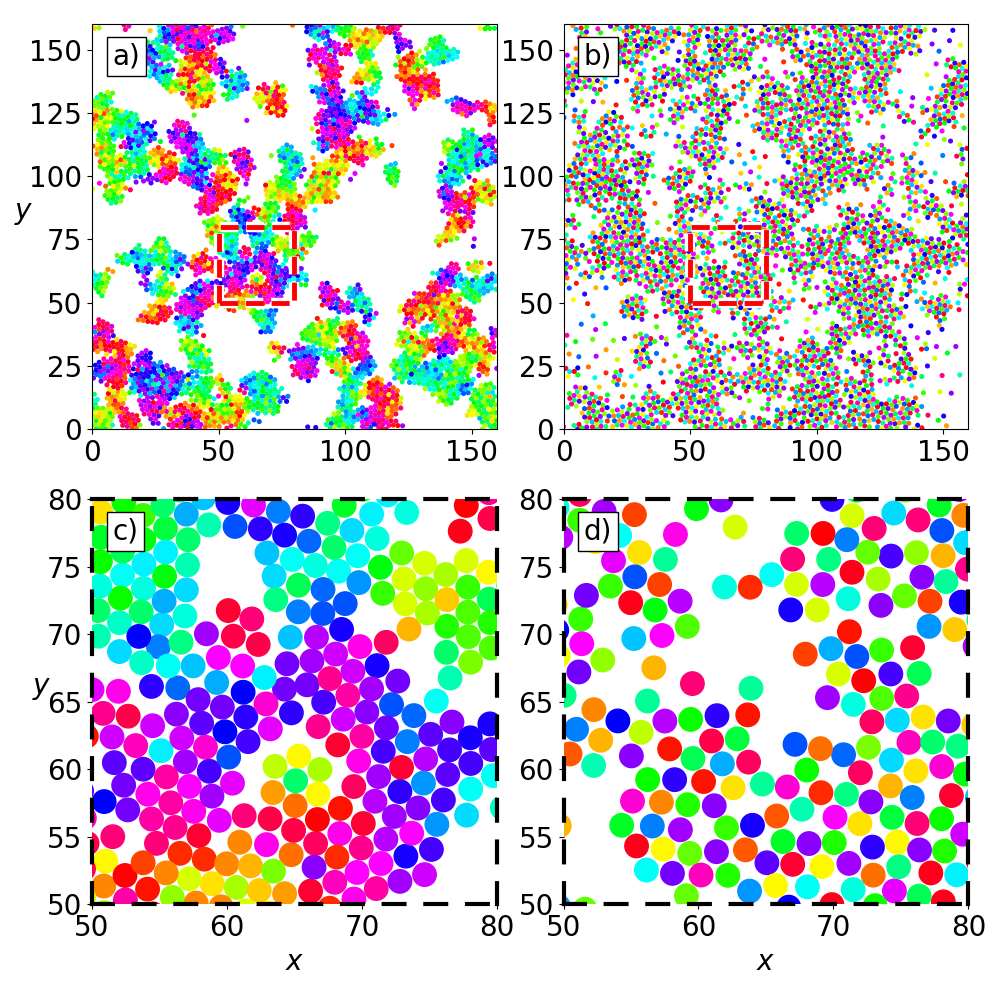}
\caption{
Illustrations of the disk locations (circles) and internal phases
(indicated by color) for the phases I$_b$ and V$_b$
in Fig.~\ref{fig:2}
for a system with $\rho=0.49$ and $\tau=1.5 \times 10^{5}$.
(a) Region I$_b$ or the arrested gel state at $J = 1.5$ and $K = 0$,
where the system separates into clusters with
the same internal phase but cannot fully coarsen due to the
repulsive interactions between disks with opposite phases.
(b) Region V$_b$, the fluctuating clump state
with local anti-synchronized phase ordering, at $J = 1.5$ and $K =-0.01$.
(c) A blowup of the red dashed box for region I$_b$ in panel (a). 
(d) A blowup of the red dashed box for region V$_b$ in panel (b).
}
\label{fig:4}
\end{figure}

In Fig.~\ref{fig:2}, we plot a dynamic phase diagram
for samples with $\rho=0.49$ and $\tau=1.5 \times 10^5$
as a function of $K$ versus $J$,
where we highlight the seven regimes of behavior.
For this choice of $\tau$, an isolated disk undergoing no collisions
would travel a distance $d_{\rm free}$ ranging from
$d_{\rm free}=150$ to $d_{\rm free}=300$, comparable to the system size.
When $K = 0$ and $J = 0$ the system forms an ordinary MIPS state,
which we designate as region II.
This MIPS phase extends along the $J=0$ axis for all values of $K$, since
the swarmalator interaction has no effect on the disk motion
when $J=0$.
The MIPS state persists for small values of $|J|$ when the swarmalator force
remains too small to perturb the motion significantly.
In region I, where
$J > 0$ and $K > 0$,
all of the disks evolve to have the
same internal phase $\phi_{\rm bulk}$, the swarmalator force
becomes attractive,
and a gas-free gel state emerges, as illustrated in Fig.~\ref{fig:3}(a).
We find the same gel state
for most choices of $\tau$ and $\rho$,
but as $\tau$ increases, the gel forms more rapidly.
The value of $\phi_{\rm bulk}$ differs depending on the initial conditions,
but all of the disks always reach $\phi_i=\phi_{\rm bulk}$
after a transient time that becomes longer as $K$ becomes smaller.
At $K=0$, for small but positive $J$ we find
region I$_b$
where there is no global ordering of $\phi_i$.
Instead, small, gel-like clusters of coherent phase form with no
surrounding free gas state,
as shown in Fig.~\ref{fig:4}(a,c).
In both regions I and I$_b$, the gel clusters slowly drift through the
system as a result of the disk activity.

For $K>0$ and $J<0$ in Fig.~\ref{fig:2}, we find region III
where the disks synchronize to a single global internal phase value
$\phi_{\rm bulk}$
but repel each other.
Here a labyrinthine pattern appears,
as illustrated in Fig.~\ref{fig:3}(b).
Due to the activity, region III
has some characteristics of a fluctuating fluid.
The disks
do not form a crystal because the
repulsive swarmalator force drops to zero
when the particles touch ($r_{ij} = 2 R_{d}$)
and is maximal when $r_{ij}=4 R_d$.
The labyrinthine structures are similar to
the labyrinthine and cluster phases found in
systems with multiple length scale repulsive potentials
\cite{Jagla98,Malescio03}.
In our case, there is a core repulsive force from the
elastic repulsion extending out to $r_{ij}=2R_d$ that is surrounded
by a softer intermediate range
swarmalator repulsive force from $2 R_{d} < r_{ij} \leq 4 R_{d}$.
In non-active systems where this type of two-step repulsion is present,
a range of additional crystalline and stripe-like phases
appear for varied densities \cite{Malescio03,Glaser07};
however, the activity in our system prevents the formation
of such higher-order structures.

For $ K < 0$ and $J < 0$ we observe a frustrated MIPS state that we denote
region IV in Fig.~\ref{fig:2}. This consists of a high-density solid with
stripe-like internal phase ordering coexisting
with a fluid that contains locally phase ordered patches,
as shown in Fig.~\ref{fig:3}(c). A blow up
in Fig.~\ref{fig:3}(f) of one of the dense regions
indicates that disks with a given value of $\phi_i$
form stripes that are interleaved between disks with the opposite phase.
A considerable number of lattice defects are present in the dense patches
and the internal phases are continuously changing due to the activity.
Within region IV, the disks are attempting to anti-synchronize their
internal phase with the phases of the neighboring particles.
Once this anti-synchronization is successful, the disk is attracted
to its neighbors, but disks with the same value of $\phi_i$ repel
each other.
Formation of the striped phase ordering is favored because this allows
each disk to maximize the number of nearest neighbors with
attractive interactions
while minimizing the number of neighbors
with repulsive interactions.
The internal phase ordering pattern
resembles the patterns found for frustrated spins on a hexagonal lattice
or the buckling of colloidal particles that have been packed into
slightly more than a monolayer \cite{Han08,Shokef09,Hill23}.
In the case of the colloidal system,
the colloidal particles are a little too dense to form a triangular solid,
but can maintain their triangular arrangement by buckling out of plane,
giving each particle an effective spin degree of freedom and
permitting the system to form an antiferromagnetic Ising-like state
that is frustrated on a hexagonal lattice.
The frustrated state breaks up into multiple
stripe-like domains that each have different orientations of the buckled
pattern.

Although region IV is related to a MIPS state, it has a larger number
of disks in the gas state compared to ordinary MIPS and these disks
exhibit a local phase ordering, as illustrated
in Fig.~\ref{fig:3}(e). The gas phase particles form loose pairs that have
opposite values of $\phi_i$.
This ordering occurs only very locally and the specific values of $\phi_i$
for the opposing pairs vary from one gaseous patch to the next.
Although this locally ordered liquid structure is
preferred by the system at low densities,
it is susceptible to the shear and compression produced by the activity
of the disks, so it does not remain
stable as a function of time but forms and disintegrates constantly in
order to establish
a dynamical equilibrium with the dense striped crystalline patches.

For $ J > 0$ and $K < 0$, the system forms a more stable
superlattice MIPS state, termed region V,
where the dense regions develop a hexagonal superlattice ordering,
illustrated in Fig.~\ref{fig:3}(d).
For these parameters,
disks with the same phase attract each other
and disks with the opposite phase repel each other,
but the phases of neighboring disks evolve to be
different from each other.
To minimize the competition between attractive and repulsive
interactions, the system forms the dense hexagonal superlattice
ordering shown in more detail in Fig.~\ref{fig:3}(g).
The swarmalator force decreases as $r_{ij}$ decreases and vanishes
when $r_{ij}=2R_d$, so disks that are in direct contact can avoid
being repelled by each other.
Locally the system then resembles repulsively interacting
internal phases on
a triangular lattice, and the superlattice ordering emerges
from the synchronization of the evolution of neighboring phases.

In Fig.~\ref{fig:3}(h), we show the regular MIPS state or region II
at $K = 0$ and $J = 0$, where the disks maintain the same random phases
they received
during initialization; however, since the internal phases play no role
in the dynamics, the dense region develops no phase ordering.
The regular MIPS state extends along
the $K$ axis for a band of finite width in $J$, since when $J$ is
sufficiently weak the evolution of the internal phases does not change
the disk dynamics.
We also find a window of region II for larger positive $J$ and
small but finite negative $K$;
here, the anti-synchronization is not strong enough
to alter the internal phases and produce region V or V$_b$ behavior.

In Fig.~\ref{fig:4}(a), we illustrate the configuration for
what we call region I$_b$ or the phase arrested gel,
which occurs for $K = 0$ and $J > 1$.
In this case, the system starts with the disks assigned to
random phases, but since $K=0$ these phases cannot evolve with time,
so the system can be viewed as containing active particles
that have randomized attractive and repulsive interactions with each other.
Over time, the disks segregate into clumps of
uniform phase, but formation of a single clump as in the
gel state or region I is not possible because clumps with opposite
phase repel each other.
A blow up of this state appears in Fig.~\ref{fig:4}(c).
The frustrated clump fluid or region V$_b$ that appears
for $K<0$ and $J>1.5$
is illustrated in
Fig.~\ref{fig:4}(b).
The internal phase configuration is not ordered within the solid
clumps and the clusters are less compact and show larger fluctuations
than the region II regular MIPS state.
Although there is no superlattice ordering within the cluster,
particles with the same phase generally
try not to be next to each other.
This region appears when $J$ becomes
large enough that the repulsive and attractive forces
destabilize the solid that forms in region V.

\begin{figure*}[ht]
\includegraphics[width=\textwidth]{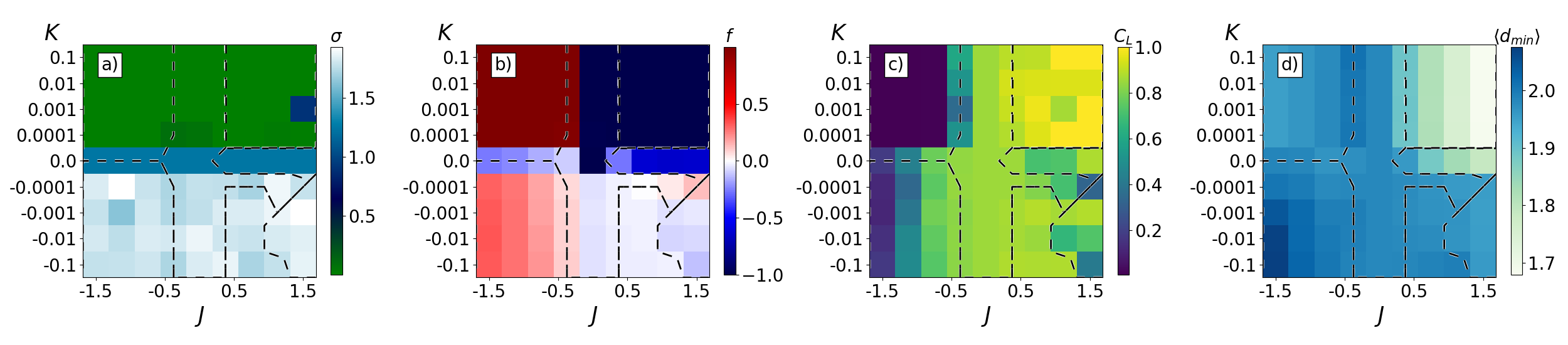}
\caption{
Illustration of different measurements that can be used to quantify the
appearances of the different regimes. Each is plotted as a heat map
as a function of $K$ vs $J$ in a system with
$\rho=0.49$ and $\tau=1.5 \times 10^5$.  
(a) Standard deviation  $\sigma$ of the circularly averaged phase
$\Bar\phi$ of the disks.
(b) The average frustration $f$.
(c) Fraction $C_L$ of disks contained in
the largest cluster present in the system.
(d) The average shortest distance $\langle d_{\rm min}\rangle$
between neighboring disks.
}
\label{fig:5}
\end{figure*}

The different regimes can be distinguished using several measures.
The circularly averaged internal phase is given by
\begin{equation}
\Bar{\phi} = \dfrac{1}{N}\sum_{i=1}^{N}\phi_i \ .
\end{equation}
This quantity is calculated every 5000 simulation time steps and
we then compute the time average $\langle \Bar \phi\rangle$ from
$M=600$ measurements.
We obtain the standard deviation
$\sigma = \sqrt{\sum_k^M(\Bar\phi_k-\langle \Bar\phi\rangle)^2/(M-1)}$,
and plot a heat map of $\sigma$ as a function of $K$ versus $J$ in
Fig.~\ref{fig:5}(a).
For $K>0$, $\sigma$ drops to zero since all of the internal
phases synchronize to a global value $\phi_{\rm bulk}$.
For $K = 0$, $\sigma$ takes on a small finite value
since the small thermal fluctuations of the
individual internal phases are no longer suppressed by synchronization,
while for $K<0$, $\sigma$ becomes large since the disks evolve to
have phases opposite from those of their neighbors.

We define the average frustration $f$ as the sum,
\begin{equation}
f = \left\langle \dfrac{1}{N}\sum_{i=1}^{N} f_i\right\rangle \ ,
\end{equation}
of the individual disk frustration values $f_i$,
\begin{equation}
f_i = -\dfrac{1}{n}\sign (J) \sign (K)\sum_{j=1}^{N} \cos(\delta\phi_{ij})\Theta(4R_d-r_{ij}) \ ,
\end{equation}
where the average is taken over time.
An individual disk with no frustration has $f_i=-1$.
Examples where frustration does not occur include disks
in region II with $J=0$ and $K=0$,
disks that have no neighbors within a radius $r_{ij}=4R_d$,
or disks for which all of the neighbors satisfy
both the $J$ and $K$ interactions such as in the gel state of region I.
In Fig.~\ref{fig:5}(b) we plot a heat map of $f$,
which ranges from $-1 \leq f \leq 1$, as a function of $K$ versus $J$.
We find $f=+1$ (maximum frustration) in region III
where the internal phases are aligned but the
swarmalator interaction forces are repulsive.
In the frustrated MIPS state of region IV,
$f$ is close to zero but negative, indicating intermediate frustration,
while in region V or the superlattice MIPS state,
the frustration is intermediate but positive.
Generally, as $f$ gets closer to $f=+1$, the
motion in the system becomes more fluid-like.

The average fraction $C_L$ of disks in the largest cluster is
defined by identifying groups of disks that are all in steric contact
with each other as members of a single cluster. This quantity is computed
using an efficient neighbor lookup table method \cite{Luding99} and is
then averaged over time.
In Fig.~\ref{fig:5}(c) we plot a heat map of $C_L$ as a function of
$K$ versus $J$.
In the active gel region I, $C_L$ is large, while in
the active labyrinthine region III,
$C_L$ is low since the repulsive
swarmalator forces break the labyrinth
structures into small disjoint clusters.
For the MIPS-like regions IV and V, there is a substantial amount of
clustering but $C_L$ is lower than its value in the standard MIPS
region III.
We observe reduced clustering in region V$_b$,
and the value of $C_L$ is one of the criteria
that we use to distinguish region V$_b$ from region V. 

In systems where pattern formation is possible, the average distance
to the closest neighbor $\langle d_{\rm min}\rangle$
can be used to identify signatures of
different patterns \cite{Reichhardt10}.
The closest neighbor of disk $i$ is at a distance
$d_{\rm min}^i=\min\{r_{ij}\}$
and we obtain
$\langle d_{\rm min}\rangle=\langle N^{-1}\sum_i^Nd_{\rm min}^i\rangle$, where
the average is taken over time.
Figure~\ref{fig:5}(d) shows a heat map of
$\langle d_{\rm min}\rangle$ plotted as a function of $K$ versus $J$.
In the active gel region I, $\langle d_{\rm min}\rangle$ is small since
all of the disks are in one giant cluster and the
swarmalator force acts as a surface tension that
further compresses the disks.
There is a linear dependence of $\langle d_{\rm min}\rangle$
on $J$ in the active labyrinthine region III 
since the increasing repulsion slightly reduces the size of the
small clusters that appear in the labyrinth formation.
In the frustrated MIPS region IV,
$\langle d_{\rm min}\rangle$ reaches its lowest value
for large $J$ and $K$ where
the stripe-ordered dense clusters illustrated in Fig.~\ref{fig:3}(f) coexist
with the maximum amount of
the low density phase-correlated state shown in Fig.~\ref{fig:3}(e).
We used the values of $\sigma$, $f$, $C_L$, and $\langle d_{\rm min}\rangle$ to
construct the dynamic phase diagram plotted in Fig.~\ref{fig:2}.

\begin{figure*}[ht]
\includegraphics[width=\textwidth]{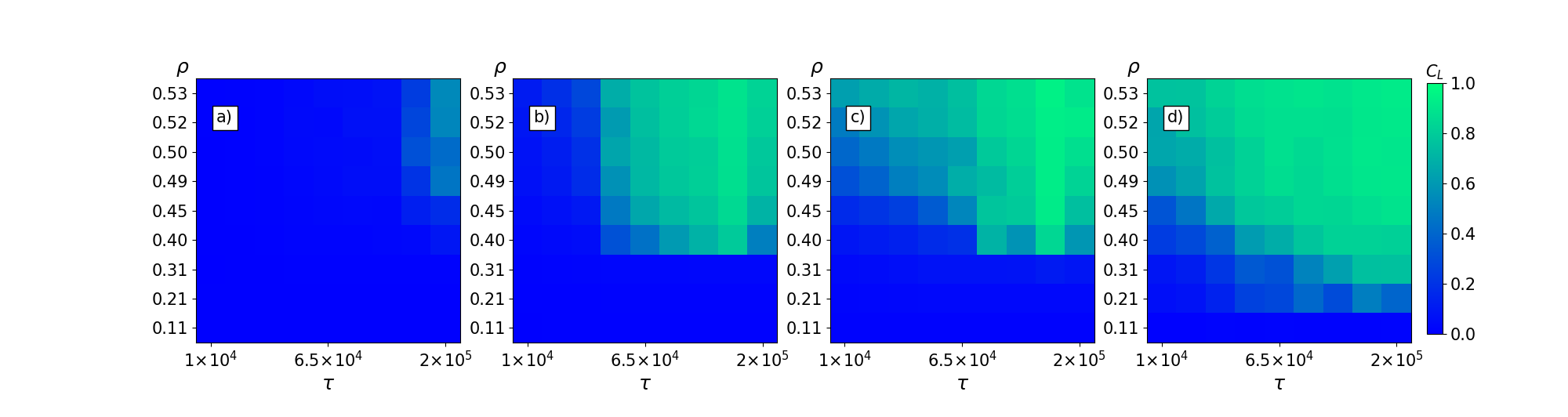}
\caption{
Behavior of the different MIPS
and cluster regimes illustrated with heat maps
of the fraction $C_L$ of disks in the largest cluster plotted as a
function of disk density $\rho$ versus run time $\tau$.
(a) The frustrated MIPS region IV at $J= -2$ and $K = -0.001$.
(b) The regular MIPS region II at $J = 0$ and $K = 0.$ 
(c) The superlattice MIPS region V at $J = 1$ and $K = -0.01$.
(d) The frustrated cluster fluid region V$_b$ at $J = 2$ and $K = -0.001$. 
}
\label{fig:6}
\end{figure*}

We next consider the impact of changing the activity $\tau$ and the density
$\rho$ on the dynamical behavior.
In general, the gel regions I and I$_b$ persist down to the lowest
values of $\tau$, but the process of gel aggregation becomes slower as
$\tau$ decreases.
The active labyrinthine region III is not modified by changes in $\tau$.
We focus on the behavior of the MIPS states in
Fig.~\ref{fig:6}, where we plot heat maps of $C_L$ as a function of
$\rho$ versus $\tau$.
In the frustrated MIPS region IV, shown in Fig.~\ref{fig:6}(a)
at $J=-2$ and $K=-0.001$,
MIPS only occurs for the highest densities and activities,
indicating that the frustration interferes with the emergence of the
MIPS.
For the standard MIPS region II with $J=0$ and $K=0$, plotted
in Fig.~\ref{fig:6}(b),
MIPS extends down to a minimum density
of $\rho=0.4$ and a minimum run time of $\tau = 5\times10^4$,
spanning a much larger window of $\rho$ and $\tau$ than the frustrated MIPS
region IV.
MIPS is enhanced for the superlattice MIPS
region V in Fig.~\ref{fig:6}(c),
where $J = 1$ and $K = -0.01$.
The superlattice ordering
illustrated in Fig.~\ref{fig:3}(g)
stabilizes the MIPS clusters and
allows them to persist down to
lower values of $\tau$.
The greatest enhancement of $C_L$ occurs in region V$_b$, shown
in Fig.~\ref{fig:6}(d) at
$J = 2$ and $K = -0.001$.
For these parameters, the system is not far from
the arrested gel region I$_b$ and the regular MIPS region II.
Although the long range crystalline ordering of the superlattice
found in the superlattice MIPS region V is absent,
there is a longer range
correlation of the values of $\phi_i$ within the clustered
areas, as illustrated in Fig.~\ref{fig:4}(d).
Neighboring disks have different phases $\phi_{i}$, but next-neighbor disks
at a distance
$r_{ij} = 4R_{d}$ have matching values of $\phi_i$ and tend to
stabilize the cluster. 
This allows the disks in the gas phase to
adhere more readily to the surface of the largest cluster
since it is not necessary for them to attach at the precise
location that would be required if long range ordering were
present. As a result, the window of large $C_L$ is maximized
for the frustrated cluster fluid region V$_b$.

\section{Discussion and Summary}

Our active matter swarmalator model demonstrates that ordinary MIPS can be considered as a special case of a larger class of possible active phase separating behaviors. The internal phase degree of freedom of each disk couples to the swarmalator rules and produces additional repulsive and attractive interactions between disks that can compete with or facilitate MIPS formation.
The swarmalator model generates a wide variety of active phases,
including an aggregating gel where activity speeds up aggregation,
a phase arrested active gel, regular MIPS,
frustrated MIPS with stripe phase ordering,
a superlattice phase ordered MIPS,
and several correlated active fluids with local phase ordering.
Future investigations could address
whether there are additional dynamics within the phase-ordered solids,
such as periodic oscillations or propagating phase waves.
Additional dynamic behaviors could arise
if a periodic external driving force
were coupled to the internal phase of the disks.
Previously considered modifications to generic swarmalator models,
such as temperature \cite{Sar22,Hong23} or chirality \cite{Ceron23b},
could also be introduced to the active disk system.
In addition, instead of a continuously variable internal phase,
the internal phase could be limited to discrete values in order to draw
parallels with Ising or other spin systems.
Our results are relevant to active colloidal systems with
competing interactions, biological systems with
particle-specific interactions, and robotic swarms.
This work provides a model for interactions between frustrated systems
and the MIPS state, and shows that in some cases, the frustration works
against formation of MIPS, but under other conditions, the frustration
can instead enhance MIPS.

\begin{acknowledgments}
	
This work was supported by the US Department of Energy through
the Los Alamos National Laboratory.  Los Alamos National Laboratory is
operated by Triad National Security, LLC, for the National Nuclear Security
Administration of the U. S. Department of Energy (Contract No. 892333218NCA000001).
AL was supported by a grant of the Romanian Ministry of Education
and Research, CNCS - UEFISCDI, project number
PN-III-P4-ID-PCE-2020-1301, within PNCDI III.
\end{acknowledgments}

\bibliography{mybib}

\end{document}